\documentclass[10pt,conference,hidelinks]{IEEEtran}\IEEEoverridecommandlockouts
\usepackage{epsfig,graphicx,subfigure,psfrag,amsmath,cases}
\usepackage{latexsym,amssymb,amsmath,epsfig,subfigure,algorithm,mathtools}
\usepackage{algorithmic}
\usepackage{color}
\usepackage{url}
\usepackage{scrtime}
\usepackage{cite}
\usepackage{epstopdf}
\usepackage{subfigure}
\usepackage{bbding}
\usepackage{multicol}

\author{Zhiqiang Wei, Min Qiu, Derrick Wing Kwan Ng, and Jinhong Yuan\\
School of Electrical Engineering and Telecommunications, University of New South Wales\\
Email: zhiqiang.wei@unsw.edu.au; min.qiu@unsw.edu.au; w.k.ng@unsw.edu.au; j.yuan@unsw.edu.au\vspace{-7mm}
\thanks{This work was partially supported by the Australia Research Council Discovery Project DP190101363 and Linkage Projects (LP 160100708 and LP170101196).
D. W. K. Ng is supported by the Australian Research Council's Discovery Early Career Researcher Award (DE170100137).
}}

\title{A Two-Stage Beam Alignment Framework for Hybrid MmWave Distributed Antenna Systems}

\newtheorem{T-Prob}{Transformed Problem}

\DeclareMathOperator{\maxo}{maximize}

\newcommand{\abs}[1]{\lvert#1\rvert}

\begin{document}
\maketitle
\begin{abstract}
In this paper, we investigate the  beam alignment problem in millimeter-wave (mmWave) distributed antenna systems where a home base station communicates with multiple users through a number of distributed remote radio units (RRUs).
Specifically, a two-stage schedule-and-align (TSSA) scheme is proposed to facilitate efficient communications.
In the first stage, a coarse beam scanning over the entire angular space is performed while beam indices and the corresponding peak-to-background ratios of the received power-angle-spectrum are obtained from users' feedback.
Then, by exploiting the user feedback, an efficient user scheduling algorithm is developed to improve the system spectral efficiency and to reduce the system misalignment probability.
Next, the second stage of beam search is performed by each RRU with reconfigured search angles, search steps, and power levels to obtain a refined beam alignment.
Simulation results show that the proposed TSSA scheme can significantly outperform the conventional one-stage method in both centralized and distributed mmWave systems in terms of beam alignment accuracy and spectral efficiency.
\end{abstract}

\section{Introduction}
The explosive growth of data-hungry mobile applications has imposed a stringent requirement of ultra-high data rate for future wireless communications\cite{wong2017key,wei2019performance}.
Thanks to the abundant available bandwidth in millimeter-wave (mmWave) frequency bands, mmWave communication technology \cite{Rappaport2013} has been recognized as a promising candidate technology to substantially improve the system capacity\cite{Rappaport2013,zhao2017multiuser}.
Unfortunately, the inherent severe path loss and the vulnerability to blockages in mmWave frequency bands are the fundamental obstacles to be tackled for achieving the promising data rates of mmWave communications\cite{Rappaport2013}.
Exploiting the short wavelength in mmWave frequency bands allows the dense packing of a large-scale antenna array in compact wireless transceivers.
This offers the possibility of forming highly directional information-carrying beams to provide a significant array gain for compensating the propagation loss.
To strike a balance between hardware complexity and performance in mmWave systems, instead of equipping each antenna with a dedicated radio frequency (RF) chain, hybrid architectures \cite{zhao2017multiuser} have been proposed.
In practice, realizing efficient beamforming in hybrid architectures requires the employment of inexpensive phase shifters (PS).
However, the establishment of highly directional transmission links in mmWave systems relies on accurate beam alignment, e.g. \cite{Alkhateeb2014,LiuChunShan2017,Li2018explore}, between transmitters and receivers during the initial access phase.
This task is often quite challenging due to the low pre-beamforming signal-to-noise ratio (SNR).
To unlock the potential of mmWave systems, this paper mainly focuses on the design of a practical scheme for handling the fundamental issue of initial beam alignment in mmWave communications.

Apart from the serious path loss, mmWave communications suffer from severe blockage effects\cite{Rappaport2013,Andrews2017}.
In particular, the line-of-sight (LOS) component of mmWave channels might be blocked by some large-size obstacles along the mmWave propagation channels, such as tall buildings, which dramatically degrades the system performance.
To provide the \textit{LOS diversity} against the blockages, distributed antenna systems (DAS), with multiple remote radio units (RRUs) assisting the home base station (BS) via high-bandwidth low-latency dedicated connection\cite{HeathDistributed}, were recently proposed to be applied to mmWave communication systems\cite{BaiLinDASs,zhao2019distributed}.
Intuitively, in mmWave DAS, the probability that all LOS paths between a user to all the distributed RRUs are completely blocked is significantly lower than that of a centralized system setting.
However, fully exploiting the LOS diversity of mmWave DAS requires an accurate beam alignment between RRUs and users.

Recently, the beam alignment problem has been intensively studied in mmWave centralized systems\cite{Alkhateeb2014,LiuChunShan2017,Li2018explore}.
To elaborate a little further, the authors in \cite{LiuChunShan2017} compared the beam alignment method based on exhaustive search and hierarchical search \cite{Alkhateeb2014}.
They demonstrated that the exhaustive search always outperforms the hierarchical method in terms of both misalignment probability and the worst case spectral efficiency, especially in the low SNR regime.
However, the exhaustive search incurs a larger time-overhead compared to that of the hierarchical search.
Furthermore, the authors in \cite{Li2018explore} proposed an optimized two-stage search algorithm for beam alignment via employing judiciously designed energy budget allocation during the two-stage beam scanning.
However, the existing beam alignment methods were designed for mmWave centralized systems\cite{Alkhateeb2014,LiuChunShan2017,Li2018explore}, which cannot be directly applied to mmWave DAS.
To the best of the authors' knowledge, the effective beam alignment scheme designs for mmWave distributed antenna communication systems have not been reported in the open literature yet.

This paper proposes a two-stage beam scheduling and alignment scheme via exploiting the features of mmWave DAS.
In particular, the mmWave DAS inherently provides the LOS diversity, which is robust against beam misalignment since misalignment only occurs if a user was misaligned with respect to (w.r.t.) all the distributed RRUs.
In addition, each RRU can only serve a limited number of users due to its small number of equipped RF chains.
If the DAS has some prior information about users' angle-of-departure (AOD), upon effective user scheduling, each RRU only needs to search for a smaller angular domain of its scheduled users, rather than searching the entire angular space.
The reduced search space provides a better use of energy for beam training to discover more users and to refine beam alignment of the discovered users.
To this end, a two-stage schedule-and-align (TSSA) scheme for mmWave DAS is proposed.
%
The first stage includes a coarse beam alignment performed at all distributed RRUs and each user sends its feedback of the best beam index and its detection confidence level to the home BS.
Based on the user feedback, the home BS performs scheduling to associate users with RRUs and reconfigures the beam search space and power level.
After receiving the beam scanning configurations from the home BS, each RRU refines the beam alignment to its scheduled user by focusing only a reduced search space in the second stage.
Simulation results demonstrate the superior performance of our proposed scheme in terms of beam misalignment probability and spectral efficiency, compared to the conventional one-stage exhaustive search-based approach in both the centralized \cite{LiuChunShan2017} and distributed systems.

%
%
%
%
%

Notations used in this paper are as follows. Boldface capital and lower case letters are reserved for matrices and vectors, respectively. $\mathbb{C}^{M\times N}$ denotes the set of all $M\times N$ matrices with complex entries; ${\left( \cdot \right)^{\mathrm{T}}}$ denotes the transpose of a vector or a matrix and ${\left( \cdot \right)^{\mathrm{H}}}$ denotes the Hermitian transpose of a vector or a matrix;
$\abs{\cdot}$ denotes the absolute value of a complex scalar
and $\left\|\cdot\right\|$ denotes the Euclidean norm of a vector.
The circularly symmetric complex Gaussian distribution with mean $\mu$ and variance $\sigma^2$ is denoted by ${\cal CN}(\mu,\sigma^2)$.

\vspace{-2mm}
\section{System Model}
\begin{figure}[t]
\centering\vspace{-1mm}
\includegraphics[width=2.6in]{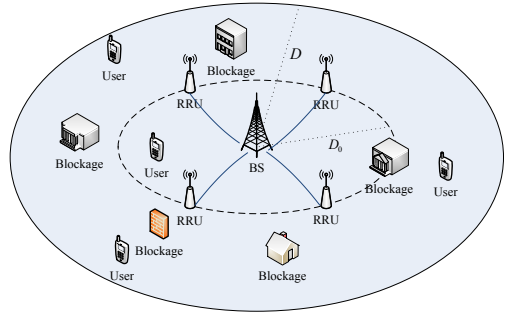}\vspace{-2mm}
\caption{The system model for a downlink mmWave DAS.}\vspace{-2mm}
\label{fig:system model}
\end{figure}

In this work, we consider a downlink mmWave DAS with $N$ RRUs and $K$ users in a micro cell, as shown in Fig. \ref{fig:system model}.
We assume that $N=K$, while each RRU and user have $M$ transmit antennas and one receive antenna, respectively.
%
Each user and RRU are equipped with a single RF chain\footnote{As a first attempt to investigate the beam alignment in mmWave DAS, we start with a simple system model consisting of single-antenna users and single-RF chain RRUs. In our journal version, we will extend the proposed beam alignment scheme to the case with multi-antenna user equipments and multi-RF chain RRUs.}.
Additionally, we assume that the AODs of each user at each RRU are independent with each other.
All the RRUs are connected to the home BS through optical fibre cables with unlimited fronthaul capacities\footnote{We note that the limited-capacity fronthaul may lead to a performance degradation of the proposed scheme. In this case, a distributed implementation with limited signalling and information exchange overhead is preferred, which will be considered in our future work.}.
The home BS communicates to all the users through RRUs in mmWave frequency band to provide a high data rate communication, while low-frequency links between the home BS and users are reserved for control signal and user feedback.
Therefore, we can assume reliable links from users to the home BS are available for information feedback during beam alignment.

\begin{figure}[t]
\centering\vspace{-1mm}
\includegraphics[width=1.7in]{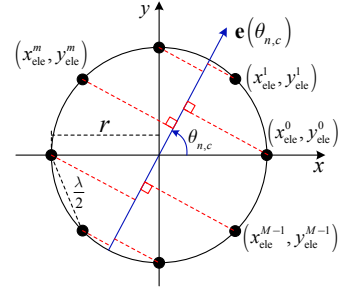}\vspace{-2mm}
\caption{The UCA structure at each RRU. Each dot denotes an antenna.}\vspace{-7mm}
\label{fig:array model}
\end{figure}

Here, the RRUs are uniformly distributed over a circle (dash line in Fig. \ref{fig:system model}) whose radius $D_0$ is half of the cell radius $D$.
We assume that each RRU is equipped with an uniform circular array (UCA)\cite{van2002optimum}, as shown in Fig. \ref{fig:array model},  where $\lambda$ denotes the wavelength at the carrier frequency.
The radius of the UCA is set as $r = \frac{\lambda}{4\sin\left(\frac{\pi}{M}\right)}$ to satisfy the antenna spacing of $\frac{\lambda}{2}$.
In fact, UCA is a general array structure which can be easily extended to an arbitrary antenna array\cite{van2002optimum}.
Besides, compared to the uniformly linear array (ULA), it can avoid the left-right ambiguity problem\cite{van2002optimum} to cover the entire angular space to facilitate the exploitation of LOS diversity.
To perform the beam alignment between RRUs and users, the $n$-th RRU first selects one pilot sequence with length of $T$ symbols, i.e., $\mathbf{s}_{n}\in \mathbb{C}^{T\times 1}$ satisfying $\|\mathbf{s}_{n}\|^2 = T$, from an orthogonal pilot set $\mathcal{C}_{\rm{P}} = \{\mathbf{s}_{n},n = 1,\ldots,N\}$, $T \ge N$.
Then, the $n$-th RRU encodes pilot sequence $\mathbf{s}_{n}$ to a codeword selected from its beamforming codebook $\mathcal{C}_{{\rm{B}},n} = \{\mathbf{w}_{n}\left(\theta_{n,c}\right),c = 0,\ldots, C-1 \}$ and broadcast it to all the users, where the codebook size $C$ equals the total steps for beam scanning.
Here, $\mathbf{w}_{n}\left(\theta_{n,c}\right) \in \mathbb{C}^{M \times 1}$ represents the array response vector of the $n$-th RRU at the AOD $\theta_{n,c}$ with $\|\mathbf{w}_{n}\left(\theta_{n,c}\right)\|^2 = 1$, for $n = 1,\ldots,N$.
The beamforming codebook $\mathcal{C}_{{\rm{B}},n}$ is determined by its beam search space, i.e., $\theta_{n,c}$, $\forall c$, which further depends on its received beam scanning configurations, with beam scanning \emph{center} $\overline{\theta}_{n}$, search \emph{range} $\Theta_n \sim [0,\pi]$ and search \emph{step} $\Delta_n = \frac{2\Theta_n}{C}$.
The detailed beam scanning configurations will be discussed in the next section.
To generate an analog beam towards $\theta_{n,c}$ at the $n$-th RRU, the analog beamformer $\mathbf{w}_{n}\left(\theta_{n,c}\right)$ for its UCA \cite{van2002optimum} is given by
\vspace{-2mm}
\begin{equation}\label{AnalogBeamformer}
\mathbf{w}_{n}\left(\theta_{n,c}\right) = \frac{1}{{\sqrt {{M}} }}e^{\left( {j2\pi \frac{{\left[ {{{\bf{x}}_{{\rm{ele}}}}\cos \left( {{\theta _{n,c}}} \right) + {{\bf{y}}_{{\rm{ele}}}}\sin \left( {{\theta _{n,c}}} \right)} \right]}}{\lambda }} \right)},\vspace{-3mm}
\end{equation}
where ${{\bf{x}}_{{\rm{ele}}}} = {\left[ {r,r\cos \left( {\frac{{2\pi }}{{{M}}}} \right), \ldots ,r\cos \left( {\frac{{2\pi \left( {{M} - 1} \right)}}{{{M}}}} \right)} \right]^{\rm{T}}}$ and ${{\bf{y}}_{{\rm{ele}}}} = {\left[ {0,r\sin \left( {\frac{{2\pi }}{{{M}}}} \right), \ldots ,r\sin \left( {\frac{{2\pi \left( {{M} - 1} \right)}}{{{M}}}} \right)} \right]^{\rm{T}}}$ denote the horizonal and vertical antenna coordinations w.r.t. the center of the UCA, respectively.
Note that the analog beamformer in \eqref{AnalogBeamformer} is obtained via projecting all the antennas of UCA to the unit vector ${\bf{e}}\left( {{\theta _{n,c}}} \right) = {\left[ {\cos \left( {{\theta _{n,c}}} \right),\sin \left( {{\theta _{n,c}}} \right)} \right]^{\rm{T}}}$ at the interested AOD $\theta_{n,c}$, as depicted in Fig. \ref{fig:array model}.

During the $c$-th step of beam scanning, the received sequence $\mathbf{y}^{(c)}_{k} \in \mathbb{C}^{1 \times T}$ at the $k$-th user is given by
\vspace{-2mm}
\begin{equation}
\mathbf{y}^{(c)}_{k} = \sum\nolimits_{n=1}^{N}\sqrt{p_n}\mathbf{h}^{\rm{H}}_{n,k}\mathbf{w}_{n}\left(\theta_{n,c}\right)\mathbf{s}^{\rm{H}}_n+\mathbf{z}_{k},\vspace{-1mm}
\end{equation}
where $\mathbf{h}_{n,k} \in \mathbb{C}^{M \times 1}$ is the downlink channel vector from the $n$-th RRU to the $k$-th user and is assumed to be a constant for the whole transmission frame; and $\mathbf{z}_{k} \in \mathbb{C}^{1 \times T}$ is the additive white Gaussian noise (AWGN) with independent and identically distributed (i.i.d.) entries $\sim \mathcal{CN}(0,\sigma_k^2)$.
Denoting the total power budget for beam alignment of DAS as $p_{\rm{sum}}$, we can allocate a power $p_n$ to the $n$-th RRU for pilot transmission with $\sum_{n=1}^{N} p_n \le p_{\rm{sum}}$.
Following the widely adopted Saleh-Valenzuela model for mmWave systems \cite{wei2018multibeam,WeiBeamWidthControl}, the channel between the $n$-th RRU and the $k$-th user is given by
\vspace{-2mm}
\begin{equation}\label{eq:sys2}
\mathbf{h}_{n,k} = \underbrace{\beta_{n,k}\alpha_{n,k,0}\mathbf{h}_{n,k,0}}_{\rm{LOS}}+\underbrace{\sum\nolimits_{l=1}^{L}\alpha_{n,k,l}\mathbf{h}_{n,k,l}}_{\rm{NLOS}},\vspace{-2mm}
\end{equation}
where $\beta_{n,k} \in \{0,1\}$ captures the blockage effect in mmWave channels, e.g. $\beta_{n,k} = 0$ means that there is no LOS path from the $n$-th RRU to the $k$-th user and $\beta_{n,k} = 1$ otherwise.
A probabilistic model \cite{Andrews2017} is adopted to characterize the probability of LOS blockage:
\vspace{-2mm}
\begin{equation}\label{eq:LOSModel}
\mathbb{P}(\beta_{n,k} = 0) = 1 - \mathbb{P}(\beta_{n,k} = 1) =  1- e^{-\frac{d_{n,k}}{\varrho}},\vspace{-2mm}
\end{equation}
where $d_{n,k}$ is the distance between the $n$-th RRU and the $k$-th user and $\varrho$ is a constant which depends on the building density in the considered area.
In \eqref{eq:sys2}, the vector $\mathbf{h}_{n,k,l}\in \mathbb{C}^{M \times 1}$ is the channel vector for the $l$-th path between the $n$-th RRU and the $k$-th user, $\forall l = {0,1,\ldots,L}$, where $l = 0$ denotes a LOS path and $l > 0$ denotes a non-line-of-sight (NLOS) path.
In addition, $\alpha_{n,k,l} \in \mathbb{C}$ is the corresponding $l$-th path gain coefficient between the $n$-th RRU and the $k$-th user.
We assume that there are $L$ NLOS paths between each pair of user and RRU.

Here, we define the user misalignment probability and system misalignment probability, which are different from the conventional definitions in centralized BS single-user systems, e.g. \cite{LiuChunShan2017,Li2018explore}.
In particular, a misalignment event happens between the $k$-th user and the $n$-th RRU if ${\hat c_{n,k}} \ne {c^*_{n,k}}$, where ${\hat c_{n,k}}$ denotes the detected beam index of the $n$-th RRU at the $k$-th user and $c^*_{n,k}$ denotes the corresponding ground truth value.
A user is misaligned only if it is misaligned w.r.t. all the distributed RRUs.
As a result, we can define the user misalignment probability as $\mathrm{P}^{\rm{mis}}_k = \Pi_{n=1}^{N}\mathbb{P}({\hat c_{n,k}} \ne {c^*_{n,k}})$, $\forall k$.
Furthermore, the system misalignment is defined as that at least one user in the system is misaligned, i.e.,
$\mathrm{P}^{\rm{mis}}_{\rm{sys}} = 1 - \Pi_{k=1}^{K} (1 - \mathrm{P}_k)$.
%
%
%
%
%
%
\vspace{-2mm}
\section{Proposed TSSA Protocol}
The key idea of the proposed TSSA scheme is to obtain a coarse estimation in the first stage which provides information for the beam scanning range, step, and power dynamically in the second stage.

\begin{figure}[t]
	\centering\vspace{-2mm}
	\subfigure[Stage-one]
	{\label{fig:TSSA_StageOne} 
		\includegraphics[width=0.13\textwidth]{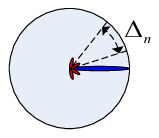}}\vspace{-1mm}
	\hspace{5mm}
	\subfigure[Stage-two]
	{\label{fig:TSSA_StageTwo} 
		\includegraphics[width=0.15\textwidth]{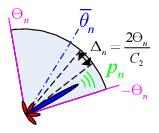}}\vspace{-1mm}
	\caption{The proposed TSSA scheme for mmWave DAS.}\vspace{-7mm}
	\label{fig:TSSA}%
\end{figure}
\vspace{-2mm}
\subsection{Coarse Beam Scanning and Information Feedback}
In the first stage, the $n$-th RRU exhaustively searches the entire angle space $\Theta_n = \pi$ with a coarse spatial resolution $C_1$, i.e., a large beam scanning step $\Delta_n = \frac{2\Theta_n}{C_1}$, via transmitting its unique orthogonal pilot sequences $\mathbf{s}_{n}$ with transmit power $p_n = \frac{p_{\rm{sum}}}{N}$ to all users, $\forall n$, as illustrated in Fig. \ref{fig:TSSA_StageOne}.
In the $c$-th step of beam scanning, the $k$-th user can extract the effective channel gain to the $n$-th RRU with the analog beamformer $\mathbf{w}_{n}\left(\theta_{n,c}\right)$ via computing the correlation between the received sequence $\mathbf{y}^{(c)}_{k}$ and the pilot sequence $\mathbf{s}_{n}$, i.e.,
\vspace{-2mm}
\begin{equation}\label{eq:PAS}
\varsigma_{n,k}(c) = \left|\mathbf{y}^{(c)}_{k}\mathbf{s}_{n}\right|^2 = \left|T \sqrt{p_n}\mathbf{h}^{\rm{H}}_{n,k}\mathbf{w}_{n}\left(\theta_{n,c}\right)+\mathbf{z}_{k}\mathbf{s}_{n}\right|^2,\vspace{-1mm}
\end{equation}
where $\varsigma_{n,k}$ is the defined power-angle-spectrum (PAS) of the $n$-th RRU obtained at the $k$-th user as a function of beam index $c$.
Then, the $k$-th user can determine the best beam index of the $n$-th RRU, i.e.,
\vspace{-2mm}
\begin{equation}\label{eq:BestBeamIndex}
{\hat c_{n,k}} = \arg\max_{c}\varsigma_{n,k}(c).\vspace{-2mm}
\end{equation}
In addition, due to the impact of the AWGN in \eqref{eq:PAS}, the estimated best beam index ${\hat c_{n,k}}$ in \eqref{eq:BestBeamIndex} can be different from the ground truth ${c^*_{n,k}}$.
Therefore, we define a metric here to capture the confidence of the correctness of the estimated beam index, we define the peak-to-background ratio (PBR), which is given by
$\xi_{n,k} = \frac{\varsigma_{n,k}({\hat c_{n,k}})}{\sum_{c \ne {\hat c_{n,k}}} \varsigma_{n,k}(c)}$.
%
%
In general, if the PAS has a higher value of PBR, the detected beam index is a better choice to reduce the beam misalignment compared to other indices.
It is reasonable that a higher PBR can be achieved with a less noise power and vice versa.
For simplicity, this paper considers the PBR as a confidence metric without justifying its optimality w.r.t. the performance of beam misalignment.
Other confidence metrics can be considered in future works.

After the coarse beam scanning, the $k$-th user feeds back the detected beam index ${\hat c_{n,k}}$ and the quantized PBR $\xi_{n,k}$, $\forall k$, of all the distributed RRUs to the home BS through a reliable communication link at the lower frequency band.
We assume that the quantization resolution for PBR is sufficient to distinguish the different confidence levels from each user to each RRU for user scheduling.
Note that, in practical implementation, several bits of quantization on PBR are adequate for efficient user scheduling, which only introduces a light feedback traffic load and a negligible feedback delay.
\vspace{-2mm}
\subsection{User Scheduling and Beam Scanning Configuration}
The user scheduling and beam scanning configuration are performed at the home BS, based on the information feedback in the first stage.
\subsubsection{User Scheduling}
Inspired by the concept of proportional fairness \cite{LauPF2005,Zhang2017Optimal}, the user scheduling design is formulated as an optimization problem to maximize the sum-log of users' confidence which is given by:
\vspace{-2mm}
\begin{align}\label{eq:UserScheduling}
&\underset{u_{n,k},\forall n,k}{\maxo} \;\;\ \sum\nolimits_{n=1}^{N} \ln\left(\sum\nolimits_{k=1}^{K} \xi_{n,k} u_{n,k}\right)  \notag\\[-0.5mm]
\mbox{s.t.}\;\;
&\mbox{C1: } \sum\nolimits_{k = 1}^K u_{n,k} = 1,\forall n, \; \mbox{C2: } \sum\nolimits_{n = 1}^N u_{n,k} = 1,\forall k,\notag\\[-0.5mm]
&\mbox{C3: } u_{n,k}\in \{0,1\}, \forall n,k,
\end{align}
\vspace{-6mm}\par\noindent
where $u_{n,k}$ is the user scheduling variable such that $u_{n,k} = 1$ means that the $k$-th user is paired with the $n$-th RRU and $u_{n,k} = 0$ otherwise.
Constraint C1 is introduced to guarantee that each RRU serves one user and C2 is imposed to ensure that each user is associated with one RRU.
The considered sum-log of confidence in the objective function in \eqref{eq:UserScheduling} aims to schedule more users and to improve the beam alignment performance.
In particular, maximizing the sum-log of confidence is equivalent to maximizing the product of users' confidence, which can efficiently improve beam alignment performance via associating users with their trustable RRUs.
Moreover, if a user does not possess trustable beam alignments to all the distributed RRUs in the first stage, the proposed user scheduling design would consider this user with a higher priority and may associate the most trustable RRU to this user.

The formulated problem in \eqref{eq:UserScheduling} is a non-linear integer programming problem and it can be transformed to the following equivalent optimization problem:
\vspace{-1mm}
\begin{equation}\label{eq:UserSchedulingII}
\underset{u_{n,k}}{\maxo} \;\ \sum\nolimits_{n=1}^{N} \sum\nolimits_{k=1}^{K} \ln\left( \xi_{n,k} \right)u_{n,k}  \;\mbox{s.t.}\; \mbox{C1}, \mbox{C2}, \mbox{C3}.\vspace{-2mm}
\end{equation}
The equivalence between \eqref{eq:UserScheduling} and \eqref{eq:UserSchedulingII} can be easily proved via verifying that both optimization problems results in the same objective value for any feasible solution satisfying C1, C2, and C3.
The resultant problem in \eqref{eq:UserSchedulingII} is a standard integer linear programming problem which can be solved via some specific numerical solvers, such as Mosek \cite{Mosek2010} via various non-polynomial algorithms.
%

\subsubsection{Beam Scanning Configuration}
Let us denote the optimized user scheduling variable via solving \eqref{eq:UserSchedulingII} as $\hat u_{n,k}$.
The home BS allocates the power to RRUs based on the confidence level of their scheduled users $\hat{\xi}_{n} = \sum_{k=1}^{K} \xi_{n,k} \hat u_{n,k}$.
It is expected to allocate a higher power level to the RRU with a lower confidence level and vice versa.
Therefore, the fractional transmit power allocation (FTPA) scheme \cite{Saito2015} is adopted where the power allocated to the $n$-th RRU can be given by
\vspace{-2mm}
\begin{equation}\label{eq:FTPA}
	{p_{n}} = \frac{p_{\rm{sum}}\hat{\xi}_{n}^{-\nu}}{\sum_{n=1}^{N}\hat{\xi}_{n}^{-\nu}},\vspace{-1mm}
\end{equation}
where $0 \le \nu \le 1$ is the decay factor.
The larger the decay factor $\nu$, the higher the power is allocated to the RRU with a lower confidence level $\hat{\xi}_{n}$.
As a result, the adopted FTPA strategy favours the RRU with a lower confidence level $\hat{\xi}_{n}$ to discover its scheduled user.

Apart from the power allocation, we desire to configure the beam scanning range based on confidence level $\hat{\xi}_{n}$ of each RRU for a better use of limited system resources.
It is expected that a narrower beam scanning range and a finer scanning step should be allocated to the RRU with a higher confidence $\hat{\xi}_{n}$, which facilitates to achieve a higher effective channel gain.
On the other hand, if a RRU has a lower confidence level for its scheduled user, a wider beam scanning range would be allocated to this RRU.
With the aid of our proposed power allocation in \eqref{eq:FTPA}, the beam alignment performance of this RRU can be improved.
In addition, we introduce a Q-function based beam scanning range control, since it is a decreasing function w.r.t. its input and its shape can be easily controlled via adjusting its mean $\mu$ and standard deviation $\sigma$.
In particular, the beam scanning range of the $n$-th RRU is given by
\vspace{-2mm}
\begin{equation}\label{eq:RangeTheta}
{\Theta _n}\left( {{{\hat \xi }_n}} \right) = \pi Q\left( {\frac{{{{\hat \xi }_n} - \mu }}{\sigma }} \right),\vspace{-1mm}
\end{equation}
with the scanning step as ${\Delta}_{n} = \frac{2{\Theta _n}}{C_2} $ and the corresponding beam scanning center as $\overline{\theta}_{n} = \frac{2\pi}{C_1}c_{n,k}$, if $\hat u_{n,k} = 1$, where $C_2$ denotes the codebook size adopted in the second stage.
Here, the mean $\mu$ and the standard deviation $\sigma$ in the employed Q-function are introduced to shape different control functions and they can be optimized based on a certain criterion.
For instance, $\sigma = 0$ results in a step function while $\sigma = \infty$ yields a linear function.
\vspace{-2mm}
\subsection{Refined Beam Alignment and Beam Index Feedback}
All the distributed RRUs receive the beam scanning configuration information from the home BS through the optical fibre cables and begin to perform a refined beam alignment in the second stage.
In particular, in the $c$-th step of beam scanning, the $n$-th RRU generates an analog beam towards the AOD $\theta_{n,c} = \overline{\theta}_{n} - {\Theta _n} + c{\Delta}_{n}$, $c = \{0,1,\ldots,C_2\}$ and transmits the pilot sequence $\mathbf{s}_n$ with power $p_n$, as shown in Fig. \ref{fig:TSSA_StageTwo}.
At the user side, similar to the first stage, after computing the correlation between the received sequence and the pilot sequence, each user can obtain its PAS and detect the refined beam index $\tilde{c}_{n,k}$.
All the users feed back the refined beam index $\tilde{c}_{n,k}$ to the home BS through the low frequency link.
The home BS can obtain the AOD of each user to its associated RRU via $\hat \theta_{n,k} = \overline{\theta}_{n} - {\Theta _n} + \tilde{c}_{n,k}{\Delta}_{n}$, $\forall n,k$.

\section{Simulation Results}
In this section, we evaluate the performance of our proposed TSSA scheme for mmWave DAS.
Unless specified otherwise, the simulation setting is given as follows.
The number of transmit antenna at each RRU is $M = 32$.
Considering a single cell with a cell radius $D = 400$ m, there are a single home BS at the center of the cell and $N$ distributed RRUs uniformly distributed on a circle with radius $D_0 = \frac{D}{2} = 200$ m.
All the $K = N$ users are randomly and uniformly distributed in the cell.
The considered mmWave DAS operates at a carrier frequency of $28$ GHz and the system bandwidth width is $100$ MHz.
The existence of LOS path from each user to RRU follows the probabilistic model as in \eqref{eq:LOSModel} and there exists $L=3$ NLOS paths.
In addition, the adopted path loss models for LOS and NLOS paths follow Table I in \cite{AkdenizChannelMmWave}.
The total system transmit power is $p_{\mathrm{sum}} = 30$ dBm and the noise power at all the users is assumed to be identical with $\sigma_k^2 = -88$ dBm.
We consider the pilot length in the range $16 \le T \le 1024$ symbols.
The decay factor in \eqref{eq:FTPA} is $\nu = 0.5$ and $\varrho$ in \eqref{eq:LOSModel} is $\varrho = D_0$.
The mean and the standard deviation in \eqref{eq:RangeTheta} is $\mu = 0.8$ and $\sigma = 0.2$, respectively.

Two baseline schemes are considered for comparison.
For baseline scheme 1, we consider a centralized subarray architecture BS equipped with $N$ $M$-antenna subarrays.
The conventional one-stage exhaustive search (OSES) is utilized for beam alignment due to its superior performance compared to the hierarchical search approach\cite{LiuChunShan2017}.
For baseline scheme 2, we consider the same DAS system as in Fig. \ref{fig:system model} but applying the conventional OSES beam alignment method\cite{LiuChunShan2017}.
Note that for baseline scheme 2, each RRU needs to select its serving user and generate an analog beam to this user.
For simplicity, random scheduling is considered for baseline scheme 2.
We assume that the codebook size is $C = M$ for the conventional OSES beam alignment.
In addition, for our proposed TSSA scheme, we assume that $C_1 = C_2 = \frac{C}{2}$, so that the beam alignment delay is the same as that of the one-stage method\footnote{Note that, by adjusting $C_1$ and $C_2$, we may obtain a better trade-off between beam alignment delay and performance. However, in this paper, we focus on comparing the beam alignment performance in terms of misalignment probability and spectral efficiency with the same delay.}.

For all the considered schemes in this section, each RRU (or subarray) generates the analog beam based on the results of beam alignment.
Then each user transmits its unique orthogonal pilot to RRUs and the home BS performs the effective channel estimation with minimum mean square error (MMSE) channel estimation technique \cite{BigueshMMSE}.
Through the use of time division duplex (TDD) and exploiting the channel reciprocity, the BS calculates the zero-forcing (ZF) digital precoder for data multiplexing based on the estimated effective channel state information.
Considering a simple equal power allocation among users, we can obtain the individual rate.
The simulation results shown in the sequel are obtained by averaging the performance over thousands of channel realizations.

\begin{figure}[t]
	\centering\vspace{-5mm}
	\includegraphics[width=3in]{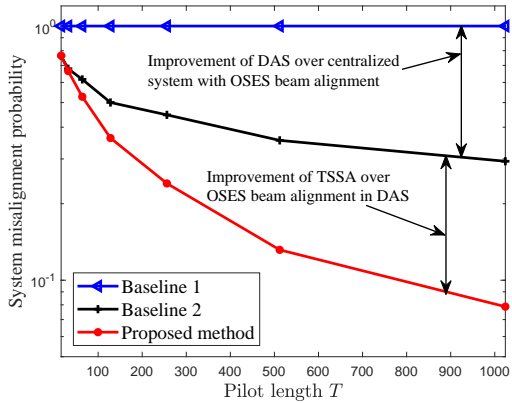}\vspace{-3mm}
	\caption{System misalignment probability versus the pilot length.}\vspace{-7mm}
	\label{fig:SysMisVersusPilotLength}
\end{figure}

Fig. \ref{fig:SysMisVersusPilotLength} illustrates the system misalignment probability versus the pilot length with $N = 8$.
Even though both schemes adopt the conventional OSES beam alignment, it can be seen that baseline 2 outperforms baseline 1 due to its inherent LOS diversity against blockages in mmWave channels.
%
In addition, we can observe that our proposed TSSA scheme can substantially reduce the system misalignment probability compared to the two baseline schemes.
Compared to the conventional OSES beam alignment, our proposed TSSA scheme can efficiently exploit the PBR feedback information and effectively schedule users to their trustable RRUs for the refined beam alignment.

\begin{figure}[t]
	\centering\vspace{-5mm}
	\includegraphics[width=3in]{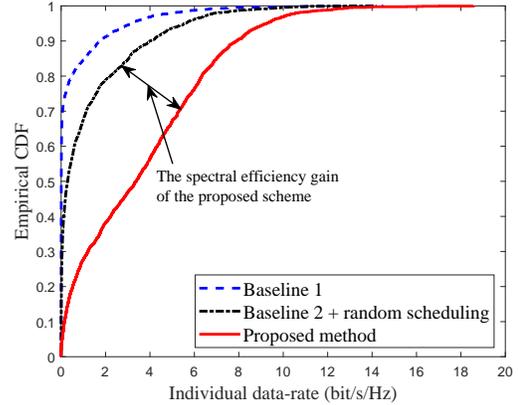}\vspace{-3mm}
	\caption{Empirical CDF of users' individual data rate. The double-sided arrow indicates the spectral efficiency gain achieved by the proposed scheme.}\vspace{-7mm}
	\label{fig:CDF}
\end{figure}

Fig. \ref{fig:CDF} illustrates the cumulative distribution function (CDF) of users' individual data rates of our proposed schemes and baseline schemes with $T = 1024$, $N=8$, and $p_{\mathrm{sum}} = 30$ dBm.
It can be seen that our proposed TSSA scheme can offer the best rate distribution, compared to both the centralized system and the distributed system with conventional OSES beam alignment.
In fact, the performance gain arises form the inherent LOS diversity of mmWave DAS, which can effectively improve the probability of the LOS existence of a user and thus improve its spectral efficiency.
On the other hand, as shown in Fig. \ref{fig:SysMisVersusPilotLength}, our proposed TSSA scheme can significantly improve the beam alignment performance and thus can achieve a larger effective channel gain, which also contributes to a higher spectral efficiency.

\section{Conclusion}
In this paper, we proposed a new two-stage schedule-and-align scheme for beam alignment in hybrid mmWave DAS to effectively exploit the LOS diversity.
Different from the conventional one-stage beam alignment scheme, our TAAS scheme first obtains the beam index and the corresponding confidence level through a coarse beam scanning within the entire angular space.
Then, user scheduling and power control are performed in order to improve the accuracy of beam alignment.
In the second stage, the beam search with reconfigured search angles and search steps is carried out to refine the beam alignment.
Simulation results demonstrated the effectiveness of the proposed scheme and the significant performance improvements of our method over the conventional schemes.

\bibliographystyle{IEEEtran}
\bibliography{NOMA}

\end{document}